\documentclass[conference]{IEEEtran}
\IEEEoverridecommandlockouts
\usepackage{cite}
\usepackage{amsmath,amssymb,amsfonts}
\usepackage{algorithmic}
\usepackage{graphicx}
\usepackage{textcomp} 
\usepackage{mathtools}
\usepackage{csquotes}
\usepackage[normalem]{ulem}

\def\BibTeX{{\rm B\kern-.05em{\sc i\kern-.025em b}\kern-.08em
    T\kern-.1667em\lower.7ex\hbox{E}\kern-.125emX}}
\usepackage{graphicx, amsmath,mathrsfs, multirow, array, siunitx, rotating, booktabs, epstopdf, subfigure, amssymb, balance, hyperref, cite,soul}
\usepackage[lined, algoruled, commentsnumbered]{algorithm2e}
\usepackage[table]{xcolor}

\newcolumntype{L}[1]{>{\raggedright\let\newline\\\arraybackslash\hspace{0pt}}m{#1}}
\newcolumntype{C}[1]{>{\centering\let\newline\\\arraybackslash\hspace{0pt}}m{#1}}
\newcolumntype{R}[1]{>{\raggedleft\let\newline\\\arraybackslash\hspace{0pt}}m{#1}}

\definecolor{morange}{rgb}{0.8,0.2,0}
\definecolor{mblue}{rgb}{0,0.1,0.8}
\definecolor{mgreen}{rgb}{0,0.8,0.1}
\definecolor{mred}{rgb}{1,0,0}

\begin{document}

\title{ A Novel Technique to Parameterize Congestion Control in 6TiSCH IIoT Networks}

\author{\IEEEauthorblockN{Kushal Chakraborty\IEEEauthorrefmark{1}, Aritra Kumar Dutta\IEEEauthorrefmark{1}, Mohammad Avesh Hussain\IEEEauthorrefmark{1} Syed Raafay Mohiuddin\IEEEauthorrefmark{1}, \\ Nikumani Choudhury\IEEEauthorrefmark{1}, Rakesh Matam\IEEEauthorrefmark{2}, Mithun Mukherjee\IEEEauthorrefmark{3}} \IEEEauthorblockA{\IEEEauthorrefmark{1} Dept. of Computer Science \& Information Systems, Birla Institute of Technology \& Science, Pilani, Hyderabad, India} \IEEEauthorblockA{\IEEEauthorrefmark{2} Dept. of Computer Science \& Engineering, Indian Institute of Information Technology Guwahati, India}
\IEEEauthorblockA{\IEEEauthorrefmark{3} Nanjing University of Information Science and Technology, Nanjing, China,}
Email: \{h20221030089, h20221030096, h20221030090, f20190382, nikumani\}@hyderabad.bits-pilani.ac.in, \\ rakesh@iiitg.ac.in, m.mukherjee@ieee.org}

\maketitle

\begin{abstract}
The Industrial Internet of Things (IIoT) refers to the use of interconnected smart devices, sensors, and other technologies to create a network of intelligent systems that can monitor and manage industrial processes. 6TiSCH (IPv6 over the Time Slotted Channel Hopping mode of IEEE 802.15.4e) as an enabling technology facilitates low-power and low-latency communication between IoT devices in industrial environments. 
%It is designed to provide reliable and deterministic communication over a wireless network, which is essential for industrial automation applications.
The Routing Protocol for Low power and lossy networks (RPL), which is used as the de-facto routing protocol for 6TiSCH networks is observed to suffer from several limitations, especially during congestion in the network. 
%Due to the inefficient congestion control mechanism of the default RPL some of the nodes have over-utilized overflowing buffers leading to high packet loss, while others have underutilized buffers. 
Therefore, there is an immediate need for some modifications to the RPL to deal with this problem. Under traffic load which keeps on changing continuously at different instants of time, the proposed mechanism aims at finding the appropriate parent for a node that can forward the packet to the destination through the least congested path with minimal packet loss. This facilitates congestion management under dynamic traffic loads. For this, a new metric for routing using the concept of exponential weighting has been proposed, which takes the number of packets present in the queue of the node into account when choosing the parent at a particular instance of time. Additionally, the paper proposes a parent selection and swapping mechanism for congested networks. Performance evaluations are carried out in order to validate the proposed work. The results show an improvement in the performance of RPL under heavy and dynamic traffic loads.    
\end{abstract}


\section{Introduction}

Over the last decade, there has been a huge rise in the use of wireless Internet of Things (IoT) devices, especially in the industrial sectors~\cite{guo2021enabling, surveyIoT, mahyoub2020efficient}. IoT has numerous applications in the industrial sector, such as predictive maintenance, supply chain management, quality control, energy management, worker safety, etc~\cite{kumbam2022lids, choudhury2018non}. 
%However, the adoption of wireless technologies in the industrial IoT has not been rapid. One of the reasons for this has been the absence of a standardized wireless protocol. 
%Many efforts were made to standardize a protocol stack that will provide a minimal profile on which required components can be bootstrapped. Some of the stacks that came up are Wireless Hart which uses TSCH as its MAC layer and uses centralized scheduling for the devices in the network. 
%It uses mesh topology and application layer commands which are defined as HART universal commands. 
%Zigbee uses CSMA for its communication but does not provide frequency hopping. It is because of this, that  ZigBee transmissions are prone to interference and threats. 
%ISA 100.11a also operates on the TSCH layer to provide reliability and security to communication. 
Due to the exponential increase in the use of constrained devices, a strategy to uniquely address these devices became an issue. In order to deal with this problem, %the Internet Engineering Task Force (IETF) 
%has standardized a protocol stack for IP-enabled constraint devices. 
IEEE 802.15.4e~\cite{802.15.4e} 6TiSCH working group has standardized the control plane and IP layer on top of the TSCH MAC layer~\cite{vilajosana2019ietf}. 
Integration of IPv6 to the TSCH network gave rise to certain issues, the proper handling of which, was extremely essential. One such challenge was addressing the arrangement of nodes in the 6TiSCH networks. To resolve this issue, an efficient network routing protocol was needed, which can route the data in the network, considering the power consumption and the limited computing capabilities of the IoT devices. In pursuit of a routing algorithm suitable for the given network, RPL was integrated with the 6TiSCH stack which was designed by the IETF Roll WG, specified in [RFC6550]. RPL is an IPv6-based routing protocol that is designed to operate in low-power and lossy networks (LLNs), and is especially suited to TiSCH-based IoT networks. RPL supports multipoint-to-point and point-to-multipoint communication, energy-efficient routing, and multicast traffic, among other characteristics that make it a good fit for LLNs. Overall, RPL is a crucial protocol for effective communication in low-power and lossy networks, which are increasingly common in the Internet of Things (IoT) and other applications.

 RPL is observed to suffer from several limitations during network congestion~\cite{righetti2023investigating, lamaazi2020comprehensive,lim2019survey}. Although IIoT devices generate small amounts of data, congestion can still occur in 6TiSCH IIoT networks due to 
 high network density, synchronization due to clock inaccuracies, scheduling conflicts, data spikes during certain events in the industrial processes, packet loss, and retransmission. Smart Grid Management is an application scenario where congestion has been observed to occur in IIoT networks. This congestion can lead to sub-optimal network efficiency and performance and, in severe cases, may even cause network failure.
 %\sout{An increase in the network size may result in congestion, leading to sub-optimal network efficiency and performance, and in severe cases may even cause network failure.} 
 Hence, it is extremely necessary to develop efficient congestion detection and control mechanisms for 6TiSCH networks. 
%From the context of practical implementation, it is never possible to decrease the congestion of a network to zero but the main objective of developing new congestion control algorithms is to make the traffic congestion as minimal as possible. 
RPL was initially designed with the primary objective of handling low and moderate traffic in the network. As a result, its performance degraded when data traffic rates in the network fluctuated or increased by a very high amount (bursty data). Therefore, in this paper, we aim to address the aforesaid issue by computing the appropriate parent to route the packets with the least congestion path. Additionally, this assists in achieving energy-efficiency by reducing packet loss and re-transmissions. 

\subsection{Motivation}
The effectiveness of congestion control depends on the network's ability to remain resilient under high-traffic conditions. Inefficient routing protocols result in network congestion. In 6TiSCH-based IoT network applications, the congestion can be avoided through the dynamic swapping of parents. The primary idea is to swap (or associate) to a less congested parent for routing the packets. However, unnecessary or excessive swapping of parents can lead to an unstable network with high swapping overhead as an association/disassociation requires the exchange of several control frames. 
Further, it may result in packet transmission failures since the child node often remains idle, waiting for new parents to be assigned most of the time. Therefore, there is a necessity to minimize the parent swapping unless the node is congested for a longer duration. In the paper, we aim to associate the child node with a less congested parent based on certain conditions discussed further.\par
%From equation (4), we are using weighted moving average. This calculation enable us to deal with two different yet very common scenarios in network topology. 
From \cite{congestioncontrolpaper}, it can be observed that when a parent node reaches a critical congestion level defined by a factor $\theta_{th}$, the child nodes are swapped. However, this threshold does not consider the parent node's previous congestion records. Additionally, a parent node may experience temporary congestion due to sporadic data bursts. In such cases, it will likely return to its normal operation mode after some time.
%It can be observed that the child nodes are swapped the moment a parent reaches a critical level of congestion set by a factor $\theta_{th}$. But it no way accounts for that parent's past records of congestion. Sometimes a parent may get congested momentarily due to bursty data. In that case that there is a high chance of the parent jumping back to its normal mode of operation after some time. 
Further, the technique considered in \cite{congestioncontrolpaper} marks the parent node from which the child nodes have been swapped due to the exceeding prescribed value of $\theta_{th}$ (critical level of congestion) as a poor candidate for parent for a $k$ amount of time. However, for sporadic data bursts, the computed mean value of $\beta$ (level of congestion at a particular node) does not change significantly, and hence, swapping a parent is not an optimal choice. 
%Our method ensures that we don't swap the parent immediately if a momentary burst of data has somehow been introduced in the channel. This is because the mean value that we will calculate as $\beta$ will remain more or less closer to a value that the parent's recent behavior has shown.\par 

\begin{figure}
    \centering
\subfigure[]{
		{\includegraphics[scale=0.415]{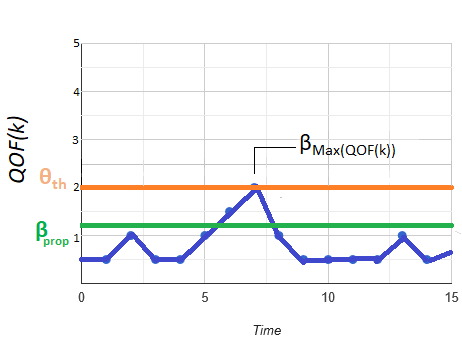}}
		\label{abrupt}}
	%\hspace*{-30pt}
	\subfigure[]{
		{\includegraphics[scale=0.415]{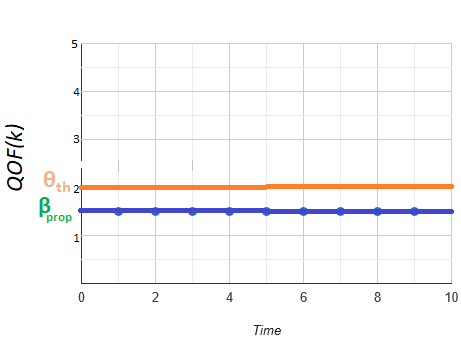}}
		\label{consistent}}
  \caption{(a) Abrupt hike in traffic and (b) consistent high traffic}
\end{figure}

We illustrate the need to reduce unnecessary swapping during sporadic data bursts in Fig.~\ref{abrupt} and Fig.~\ref{consistent}. In Fig.~\ref{abrupt}, we observe that there is a sudden data burst for a parent node resulting in a peak represented by $\beta_{\text{Max(QOF(k))}}$~\cite{congestioncontrolpaper}. This results in parent swapping as $\beta_{\text{Max(QOF(k))}}$ exceeds the $\theta_{th}$ value. However, it can be observed the traffic load reduces quickly, indicating a sub-optimal choice of parent swap. In this scenario, the average level of congestion in the considered parent node does not significantly change. This is shown by the $\beta_{\text{prop}}$ in Fig.~\ref{abrupt}. Further, Fig.~\ref{consistent} shows a parent node with consistently high traffic but below the $\theta_{th}$. This node becomes a good parent candidate for the child nodes swapped from parent nodes having traffic characteristics of Fig.~\ref{abrupt}. But, due to this swap, the candidate parent node may incur more congestion resulting in packet retransmissions and losses. This degradation in network performance could have been avoided if parent swapping is performed based on the average level of congestion in the parent, given by $\beta_{\text{prop}}$.

 % Observe in Fig.2 where we calculate $\beta$ from equation (4). Since its a slow moving average the value will remain more or less closer to its usual trend (Note: this graph doesn't depict the actual results from real data, its purpose is to illustrate what we are trying to achieve. A more detailed graph with data is discussed in Section V). Therefore $\beta$ will not change much due to sudden abrupt changes, enabling the parent to remain a better candidate. Whereas if we follow the method mentioned in \cite{congestioncontrolpaper}, we will switch to a parent mentioned in Fig.3 because its $max(QOF(k))$ is less than the one mentioned in Fig.2 although its weighted average is always higher than the current one. This will end up making the children drop more packets due to higher traffic in this new parent, whereas this would not have been the case if they had not swapped their parent. Since the older parent had jumped back to its usual performance. Further, this will make the new parent even further congested.
 % \begin{figure}[t]
 % \centering
 % \includegraphics[width=0.33 \textwidth]{scatter-plot(2).png}
 % \caption{Abrupt hike in traffic}
 % \label{abrupt}
 % \end{figure}

 % \begin{figure}[t]
 % \centering
 % \includegraphics[width=0.33\textwidth]{scatter-plot(3).png}
 % \caption{Consistent high traffic}
 % \label{consistent}
 % \end{figure}

%\begin{figure}%
 %   \centering
  %  \subfloat[\centering Abrupt hike in traffic]%%{{\includegraphics[width=5cm]{scatter-plot(2).png} }}%
  %  \qquad
   % \subfloat[\centering Consistent high traffic]{{\includegraphics[width=5cm]{scatter-plot(3).png} }}%
    %\caption{2 Figures side by side}%
    %\label{fig:example}%
%\end{figure}
\subsection{Contribution \& Organization}
 In this paper, we propose a novel technique to reduce unnecessary parent swaps and later compute an appropriate parent for routing packets. This is achieved by incorporating exponential weighting to select the parent node based on the number of packets in the node’s queue at a given instance of time. The proposed mechanism identifies and manages congestion in 6TiSCH networks.
 The contributions of the paper are summarized below.
 \begin{enumerate}
     \item Development of a parent selection and swapping mechanism for congested networks and avoiding unnecessary parent swapping during sporadic traffic data bursts.
     \item Development of a new metric for RPL-based routing using the concept of exponential weighting to compute the appropriate parent with the least congested path. This helps in managing congestion in the network.
     \item Performance comparison with existing work shows the proposed mechanism improves RPL-based several QoS measurements under heavy and dynamic traffic loads.  
     
 \end{enumerate}

The rest of the paper has been divided into the following sections: Section~\ref{related} discusses the related work in the routing domain of  6TiSCH and LLN networks. The network model considered has been discussed in Section~\ref{network-model}. Section~\ref{proposed} describes the proposed work and the algorithms used. Section~\ref{experiment} discusses in extensive detail the experimentation performed and the performance evaluations. Section~\ref{conclusion} concludes the paper.
%The rest of the paper is organized as follows. Section~\ref{related-research} describes the related works of LoRa scheduling. The network model is presented in Section~\ref{network-model}. The proposed LIDS mechanism is presented in Section~\ref{LIDS}. Subsection~\ref{rescheduling} explains the LoRa rescheduling mechanism. Experimental results are described in Section~\ref{results}. Finally, conclusions are drawn in Section~\ref{conclusion}.

\section{Related Work}
\label{related}
The rapid growth of network size caused by the implementation of IPV6 into the TSCH schedule for IIoT devices has led to a significant issue of network congestion that needs to be addressed, given that RPL was initially designed for small to moderate-size topologies.
%Due to the integration of IPV6 in the TSCH schedule in IIoT devices, the network size grew rapidly, and since RPL was proposed for small and moderate-size topologies, network congestion became a very real issue that was required to be handled. 
%There have been many efforts for addressing the issue of network congestion in RPL topologies. 
To achieve congestion control in RPL, various techniques~\cite{rfc6552, rfc6719, 7338325, 7482026, s18113838, 7374975, congestioncontrolpaper} can be employed, such as rate control, congestion notification, and congestion avoidance. Rate control involves limiting the rate at which data is sent through the network, while congestion notification involves signaling to nodes that congestion is occurring. Congestion avoidance involves adjusting the routing paths in the network to avoid congested areas. The authors in~\cite{rfc6552} relied only on objects defined by RPL and proposed a new objective function for network formation based on the rank criteria, which is merely a track of the hop count of a particular node to its DODAG sink. It also provides a mechanism to increase the rank of a particular node when the upward data transmission fails, though an increase in rank may trigger network reformation, which creates a high network overhead. Moreover, it does not discuss any method to calculate congestion or load balancing.
%which is taken care of by our proposed work.
 In~\cite{rfc6719}, an objective function that incorporates various metrics for path calculation, preferred parent selection, and parent swapping is presented. 
 %The process has two steps. First, it identifies the path with the minimum cost or Rank. Second, it only switches to this path if it has a lower cost compared to the current path by a certain threshold value. 
 For parent swapping, it uses a threshold to select a new preferred parent based on the path's cost through the current preferred parent compared to the minimum cost path. Although it prevents unnecessary swaps when an upstream transmission fails, it does not consider congestion status for either the current or preferred parent for the reallocation of child nodes. 
 %However, it does account for attributes like rank and cost for parent swapping.
The authors in~\cite{7338325} proposed a congestion control mechanism primarily focused on Queue utilization and rank i.e., the number of hops from a node $n_i$ to the sink/DODAG root node. 
%It is designed in such a way that it accounts for the queue utilization factor of its neighboring nodes and the hop distance to the border router. 
The scheme was relatively inadequate since they only considered the ratio of the number of packets in the queue to the buffer size, which would prompt a parent swap once it met a specific threshold. However, there may be instances where the data transmitted in bursts fill the queue temporarily, and the queue remains underutilized for the remainder of the network session.
%The congestion detection criteria proposed by them were rather weak as they only accounted for the ratio of the number of packets in the queue to the buffer size which will trigger a parent swap when met with a threshold. Whereas there might be a case of bursty data that may fill up the queue temporarily and for the rest of the network session the queue remains underutilized. 
~\cite{7482026} describes an objective function that addresses both low and high data transmission scenarios. 
%The authors argue that merely considering the link quality between the parent and child node does not provide a complete picture of congestion, as buffer overflow may cause packet drops and children may be unaware of empty neighboring buffers. 
The authors propose an objective function that considers link quality and buffer occupancy and a combined matrix for congestion detection with two prioritizing factors that balance suitable features for congestion detection. However, this approach has the same disadvantage with sporadic data bursts as discussed previously.  
%~\cite{s18113838}propose a congestion-aware routing protocol (CoAR) to address the issue by using the choice of an alternative parent to reduce network congestion. The suggested mechanism combines the various routing indicators to choose the optimal alternative parent node within the congestion using a multi-criteria decision-making approach. When the routing score is tied, the neighborhood index is also utilized to break ties during the parent selection procedure. CoAR is an adaptive congestion detection mechanism based on current queue occupancy and observation of both recent and historical traffic trends to identify congestion. 
%~\cite{7374975}proposed a multi-path extension of RPL, be used to offer interim multipath routing when a path is congested. The primary way a forwarding node detects congestion is by tracking the packet delivery ratio. Congestion is then reduced by offering partially disconnected multipath routing.
In~\cite{congestioncontrolpaper}, the proposed approach includes two parent selection techniques to adjust to changing network traffic loads. The level of each node's queue backlog is monitored to identify congestion, and new parent nodes are chosen following the balance of the network's load. Unnecessary parent swaps are a major issue in this technique. In view of this, we present a novel technique that minimizes unnecessary parent swaps and computes an appropriate parent node for routing packets in a congested network. 
%We compare the proposed mechanism with ~\cite{congestioncontrolpaper} as this work    %Moreover, a new routing metric is established that takes queue occupancy into account when choosing the new parent node.

\section{Network Model}
\label{network-model}
 \begin{figure}[t]
    \centering
    \includegraphics[scale=0.24]{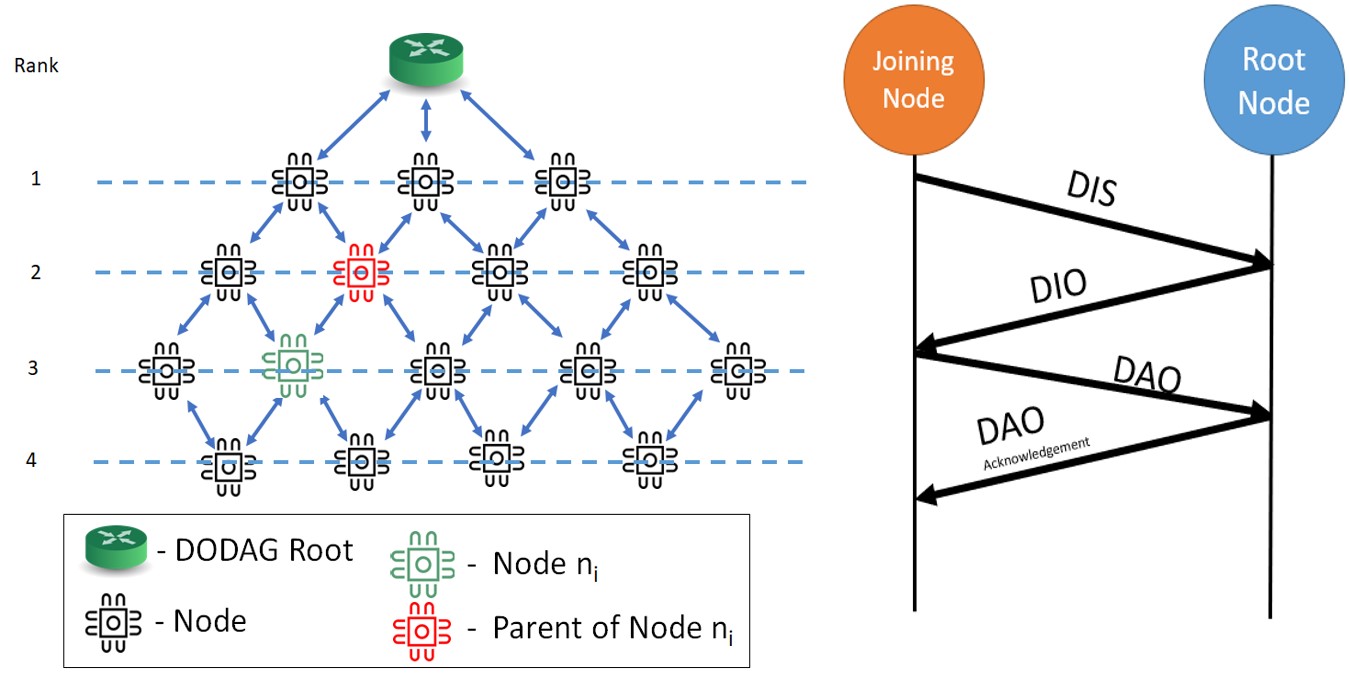}
    \caption{Destination Oriented DAG and RPL routing control messages in RPL network}
    \label{fig:model}
\end{figure}

The network model considered in this paper has been shown in Fig.~\ref{fig:model}. The network has a Destination Oriented Directed Acyclic Graph (DODAG) structure. 
%It is a special kind of Directed Acyclic Graph with a single root (or sink). 
The DODAG has a stratified routing structure that organizes nodes in a tree-like topology which has a single root node at the top, leaf nodes at the bottom, and each node has one or more parents, which are responsible for forwarding packets toward the destination. Each DODAG network topology has a version number. This same number is present in all the nodes of the network. There are two modes of operation, namely, a) \textit{Storing mode}:   wherein all the nodes contain full information of the network topology in their routing table. As a result, all the nodes have information about directly reaching the other nodes, and b) \textit{Non-Storing mode}: wherein only the border router or DODAG root has the complete information of the network topology in their routing table. All other nodes only maintain their list of parents and use this list as a route to send their packets toward the border router. In this paper, we considered the network functions in the Non-Storing Mode. A node sends a DIS control message when it wishes to join a DODAG but hears no announcement or cannot detect any DODAG network. When a node in a DODAG network receives DIS messages, it multicasts the DIO message and announces to other nodes the presence of itself and the DODAG network. Each node in the network maintains a routing table that lists the next-hop node for each destination. When a node needs to send a packet to a destination, it looks up the destination in its routing table and forwards it to the next-hop node listed. If a node doesn't have a routing table entry for a particular destination or wants to join the network, it sends a DAO message up the DODAG toward the root node. The root node then replies with a DAO-ACK message that contains information about the destination, including its rank and the next-hop node towards the destination. 
%Each intermediate node in the path between the source and destination nodes can cache this routing information, which helps to minimize the amount of signaling traffic required for routing.

\section{Proposed Work}
\label{proposed}
\begin{table}
    \centering
    \caption{Symbol Table}
    \label{tab:Hg}
    \begin{tabular}{ll}
      \toprule
        Symbols        & Meaning                                    \\
      \midrule                           
        CuP$_i$           & Current Parent node of node i \\
        CaP$_i$           & Candidate Parent of node i\\
        PS(P$_i$)       & Parent Score of node P$_i$ \\
        minPS           & minimum Parent Score  \\
        DAG             & Directed Acyclic Graph \\  
        DODAG           & Destination Oriented DAG \\
        DIS            &  DODAG Information Solicitation           \\
        DIO            &  DODAG Information Object \\
        DAO            & DODAG Advertisement Object            \\
        DAO-ACK        & DODAG Advertisement Object Acknowedgement \\
        Parent(n$_i$)    & Potential set of parent candidates \\
        ETX(n$_i$,P$_i$)   & Estimated Transmission Count from node n$_i$  to  P$_i$ \\
        TNOP(n$_i$,P$_i$) & total number of packets sent from node n$_i$ to P$_i$ \\
        TNOPSS (n$_i$,P$_i$) & total number of packets sent successfully from \\ &  node n$_i$ to P$_i$\\
        QOF (n$_i$)   & Queue Occupancy Factor of node n$_i$\\
        EWQOF & Exponential Weighted Queue Occupancy Factor \\
        HDLAC (n$_i$,P$_i$) & Hop Distance Link Assessment Criterion \\ 
        NONTP (n$_i$) & Number of non-transmitted packets of queue of n$_i$ \\
        QL (n$_i$) & Length of Queue of n$_i$ \\
        
      \bottomrule
    \end{tabular}
  \end{table}
\label{LIDS}
In this section, we present our proposed mechanism, which is based on our novel approach termed Exponential Weighted Queue Occupancy Factor (EWQOF) for identifying and managing congestion in 6TiSCH networks. Initially, the DODAG tree is formed by disseminating and receiving RPL control messages, namely DIS, DIO, DAO, and DAO-ACK. In the first instance, the parent $P_i$ is selected by a node $n_i$ from its potential set of parents $Parent(n_i)$ based on their distance from the DODAG root i.e., node $n_i$ selects the parent $P_i$ which has the minimum distance to the DODAG root. However, this congestion metric is insufficient and not useful for the nodes during the entire network lifetime due to the dynamic nature of the traffic. In the 6TiSCH network, the traffic load may change during its lifetime due to various reasons: addition of new IoT devices to the network with heterogeneous sending rate, change in packet sending rate of IoT devices during certain intervals of time, malfunction or shutdown for maintenance of certain IoT devices during certain time intervals. Our proposed work comes into action after the initial DODAG tree is formed. Next, we consider metrics that are used for the detection and prevention of congestion in the network. The first metric criterion for selecting a better potential parent is called $Rank$. The $Rank$ of a parent $p_i$ is the number of nodes between the $p_i$ and the DODAG root. Alternatively, it can be defined as the number of hops or hop distance from $p_i$ to DODAG root plus one. It is represented as $Rank(p_i)$.
 \begin{equation}
     Rank(p_i) = Hop\_distance(p_i) + 1
 \end{equation}

 The second metric criterion for selecting a parent to send the packets through the  least congested path is called $Estimated \ Transmission \ Count \  (ETX)$. $ETX$ is the ratio of the total number of packet transmissions from node $n_i$ to parent $p_i$ to the total number of successful transmissions of packets from node $n_i$ to parent $p_i$. This metric gives us a notion of the link quality connecting node $n_i$ to parent $p_i$.
 
   \begin{equation}
     ETX(n_i,p_i) = \frac{TNOP\ (n_i,P_i)}{TNOPSS\ (n_i,P_i)}
 \end{equation}

 The third metric for selecting the least congested parent from the candidate parent set is the $ Queue \ Occupancy \ Factor \ (QOF)$. It is also called the backlog factor. It is the ratio of the number of packets in the queue of $n_i$, which has not yet been transmitted to the length of the queue. 
 \begin{equation}
     QOF(n_i) = \frac{NONTP \ (n_i)}{QL \ (n_i)}
 \end{equation}

 The information about these metrics of a node is circulated to various nodes of the network through DIO messages.

\subsection{Parent Selection and Swapping Mechanism}
\label{rescheduling}
\begin{algorithm}[t!] 
    \footnotesize
    \SetAlgoLined
    \nl \textbf{Input:}  DIO messages from Parent($n_i$) \\
    \nl \textbf{Output:} Most appropriate parent $P_i^*$ with minimum Parent Score \\
    
    \nl minPS = very high value \\
    \If{$\beta(CuP_i) > \theta_{th}$}
    {
        \For{each parent $CaP_i \in$ Parent($n_i$)}
        {
            \If{$(HDLAC(n_i, CuP_i) - HDLAC(n_i, CaP_i)) > \delta_{th}$}
            {
                \nl Find out $PS (CaP_i)$ ; \\
                \If{ $minPS > PS(CaP_i)$ }
                {
                    minPS = $PS(CaP_i)$;\\
                    $P_i^* = CaP_i$;\\
                }
            }
        }
    }
    \Else {  
        \nl Do not consider that $CuP_i$ is congested yet\\
        \nl Continue to transmit packets to $CuP_i$  \\
    }  	 
    \nl \Return The Parent $P_i^*$ with minimum $PS$ 
    \caption{Proposed Parent Selection and Swapping Mechanism.}
    \label{Algo:LIDS}
\end{algorithm} 

%\begin{algorithm}[t!] 
% 	\footnotesize
% 		 \SetAlgoLined
%   \nl \textbf{Input:}  DIO messages from Parent($n_i$) \\
%   \nl \textbf{Output:} Most appropriate parent $P_i^*$ with minimum Parent Score \\
   
%  \nl minPS = very high value \\
%  \If{$\beta(CuP_i) > \theta_{th}$}
%   {
%      \For{each parent $CaP_i$ \epsilon \ Parent($n_i$)}
%      {
%         \If{$(HDLAC(n_i, CuP_i) - HDLAC(n_i, CaP_i)) > \delta_{th}$}
%         {
%            \nl Find out PS ($CaP_i$) ; \\
%            \If{ $minPS  >  PS(CaP_i)$ }
%               {
%                  minPS = $PS(CaP_i)$;\\
%                  $P_i^* = CaP_i$;\\
%               }
%            }
%          }
%    }
 %\Else {  
% \nl Do not consider that $CuP_i$ is congested yet\\
%         \nl Continue to transmit packets to $CuP_i$  \\
%  }  	 
%   \nl \Return The \ Parent \ $P_i^*$ \ with \ minimum \ %$PS$ 
% 	\caption{Proposed Parent Selection and Swapping %Mechanism.}
% 	\label{Algo:LIDS}
%  \end{algorithm} 
  
In this section, we present the proposed parent selection and swapping Algorithm~\ref{Algo:LIDS}. It describes how the parent nodes are selected from the candidate parent set to prevent congestion in the network. The different parameters used in the proposed algorithm are explained below. 

The level of congestion at a particular node ${n_i}$ is given by ${\beta (n_i)}$. It is a very essential parameter since it triggers the node ${n_i}$ to switch its parent to avert congestion.
We define ${\beta (n_i)}$ as follows:
\begin{equation}
 \begin{multlined}
   {\beta (n_i)} = {\alpha ^ k}\times QOF_{1} (n_i)+{\alpha ^ {k-1}}\times(1-\alpha)\times QOF_2 (n_i) + \\ {\alpha ^ {k-2}}\times (1-\alpha)\times QOF_3 (n_i)+...+(1-\alpha)\times QOF_k(n_i) \   \  \  \  \  \  \   
  \end{multlined}
  \label{beta}
\end{equation}
In Eq.~\eqref{beta}, we calculate the congestion level at a particular node ${n_i}$ by using the concept of QOF. Sometimes it may happen that the parent of $n_i$, i.e., $P_i$ is highly congested even though $n_i$ is not congested. In such a case, it would be an optimal choice for the child node of $n_i$ not to select the parent as $n_i$ because it would eventually lead to congestion and loss of packets. So we modify the equation of QOF as follows:
\begin{equation}
    QOF(n_i)  =  max \ \{ QOF(P_i), \frac{NONTP \ (n_i)}{QL \ (n_i)}   \}
\end{equation}
Let us assume that the current time instant is $t$, then $QOF_k (n_i)$ is the queue occupancy factor of the node $n_i$ from time instant $(t-1)-(k-1)$ to time instant $(t-1)$ . In Eq.~\eqref{beta}, $\alpha$ is called the smoothing factor. We have considered $\alpha = 0.5 $. The Eq.~\eqref{beta} calculates the congestion level of a node by taking into account its QOF for the previous $k$ time slots and giving more weightage to the QOF of the recent time slots and less weightage to the QOF of former time slots.   This concept is being applied using exponential weighting.   

The value of $k$ is determined such that $k \times \ T > I_{min} $ where $T$ is the duration of the slotframe and $I_{min}$ is the minimum interval of the Trickle Timer~\cite{fawwaz2023adaptive}. It should be noted that $k > \frac{1}{T}$. Resource limitations are common in LLN nodes, which often have restricted storage capacity. As $k$ is increased to accommodate more values of QOF, it could potentially surpass the node's storage limit, creating a constraint for the node beyond a certain point. The upper bound of $k$ is decided based on the available resources in the node.
\begin{equation}
    {\beta (P_i)} > \theta_{th}
    \label{6}
\end{equation}
Each node $n_i$ records the ${\beta (P_i)}$ of its current parent $P_i$. If the node finds that the ${\beta (P_i)}$ is greater than $\theta_{th}$, then it infers that the congestion level of its parent is high, so the loss of its transmitted packets will be high, and hence the node decides to change its parent. Thus the value of $\theta_{th}$ enables the node to decide when to switch its parent due to a high congestion level. Here we consider that the $\theta_{th} = 0.5$. The value of $\theta_{th}$ must be chosen very carefully. If the value of  $\theta_{th}$ is very high, then the congestion detection and parent swapping mechanism will get delayed. If the value is very low, then there will be unnecessary swapping of parents even with a low congestion level which will increase the power consumption and overhead in the network.

In this proposed work, the parent selection strategy combines the Rank, ETX, and level of congestion (${\beta (P_i)}$) metrics to compute node congestion and select the most appropriate parent ${P_i}^*$ with less congestion level. Using this concept, a routing metric $ Parent \ Score (PS)$  is introduced. $PS(P_i)$ is defined as follows:
\begin{equation}
    PS(P_i) \ = \ Rank(P_i) \ + \ ETX(n_i,P_i) \ + \ \eta\times  QOF (P_i)
\end{equation}
 
 where $\eta$ is a weight factor. It controls the effect of $QOF(P_i)$ in $PS$ during the parent selection process. Since $QOF$ is a fraction, so $0 \leq QOF \leq 1$. To have a significant effect of $QOF$ in $PS$, $\eta > 1$.

In addition to Eq.~\eqref{6}, there exists another criterion that needs to be satisfied to avoid unnecessary parent change. It is $Hop \ Distance \ Link \ Assessment \ Criterion (HDLAC)$. It is defined as follows:
\begin{equation}
     HDLAC(n_i,P_i) - HDLAC(n_i,P_i^*) > \delta_{th} \\
     \label{8}
\end{equation}
where,
\begin{equation}
     HDLAC(n_i,P_i) = Rank(P_i) + ETX(n_i,P_i)
\end{equation}
$\delta_{th}$ is a threshold value used to prevent rapid switching of parents or oscillations between parents in order to reduce the excessive attrition in the network. After satisfying Eq.~\eqref{6} and Eq.~\eqref{8}, the most appropriate least congested parent  ${P_i}^*$, having the minimum Parent Score, is determined and chosen as per the following expression. 
   \begin{equation}
      {P_i}^* = \min_{P_i \epsilon Parent(n_i)} \{PS(P_i)\}  
      \label{10}
   \end{equation}
%Once the congestion in the current parent is detected through equation Eq.~\eqref{6}, we select the most appropriate less congested parent using Eq.~\eqref{10}, satisfying Eq.~\eqref{8}.   
\begin{figure*}[h]
	\centering
	\subfigure[]{
		{\includegraphics[width=4.2cm,height=4.cm]{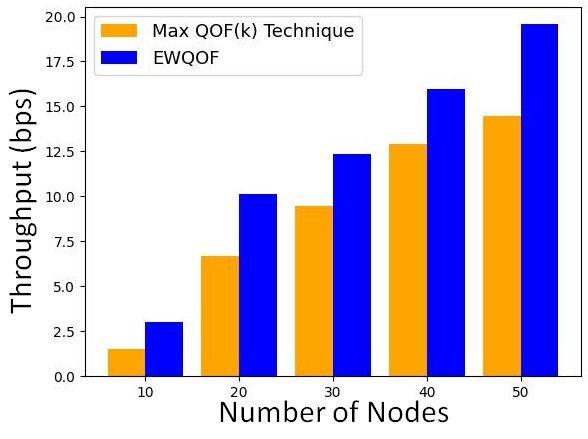}}
		\label{fig:EWQOF_thr}}
	%\hspace*{-30pt}
	\subfigure[]{
		{\includegraphics[width=4.2cm,height=4.cm]{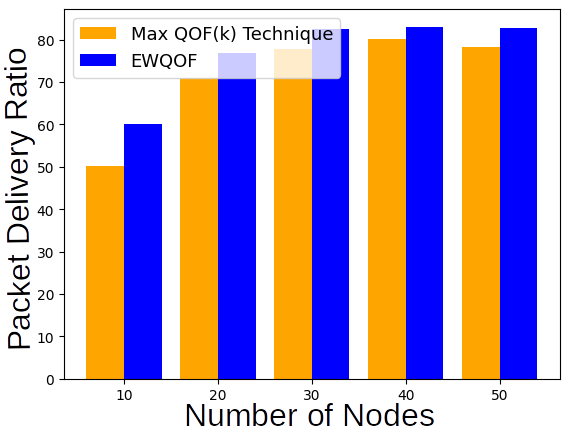}}
		\label{fig:EWQOF_pdr}}
	\subfigure[]{
		{\includegraphics[width=4.2cm,height=4.cm]{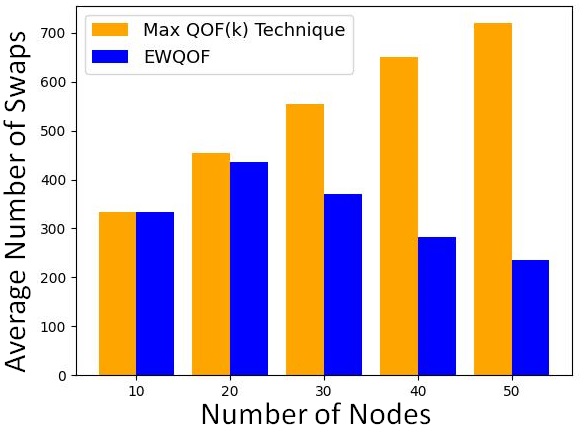}}
		\label{fig:EWQOF_swaps}}
		\subfigure[]{
		{\includegraphics[width=4.2cm,height=4.cm]{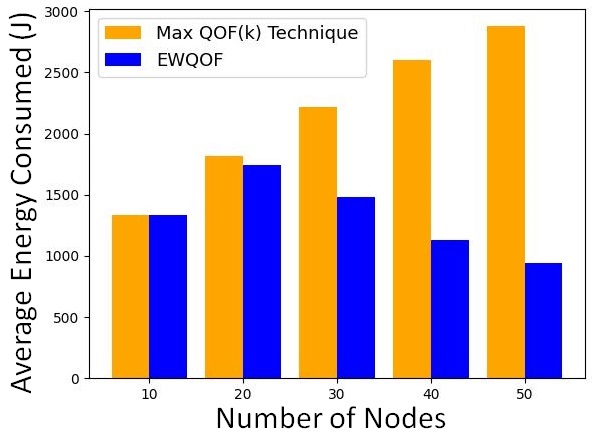}}
		\label{fig:energy}}
	\caption{(a) Throughput, (b) Packet Delivery Ratio, (c) Average Number of swaps, (d) Average Energy Consumed.}
	% \vspace{-1.25em}
\end{figure*}
\vspace{-1em}
\section{Experimental Results}
\label{experiment}
In this section, the performance of the proposed EWQOF mechanism is evaluated and compared with Max QOF scheme, which has already been proved to outperform RPL with OF0 and MHROF objective functions \cite{congestioncontrolpaper}. The performance metrics considered for the evaluation of the proposed scheme are 1) Throughput, 2) Packet Delivery Ratio, 3) Average number of Swaps, and 4) Average Energy Consumption. 
%In this section, the performance of the proposed LoRa LIDS scheme is evaluated and compared with the default mechanism and other related schemes~\cite{zorbas2019autonomous,kumari2020nodes,triantafyllou2020novel}. The performance metrics considered are: 1) transmission overhead, 2) energy consumption, 3) collision rate, and 4) throughput.

\subsection{Simulation Setup}
The network topology considered for the experiment is shown in
Fig~\ref{fig:model}. 
%The proposed method has also been evaluated by scaling the network with new devices. 
Extensive simulations were conducted in MATLAB to assess the effectiveness of the proposed approach. The parameters and settings specified in Table ~\ref{Table:SimulationParameter} were employed for the evaluations. The value of $\eta$ considered for the experiment is 0.25 \cite{congestioncontrolpaper}. The value of $\alpha$ considered for the experiment is 0.5. This value was chosen as it enables to track QOF data closely by giving a balanced weight between recent QOF data and past QOF data and provides a significant increase in Packet Delivery Ratio as compared to other values of $\alpha$.

 \begin{table}[t]
    \centering
	\caption{Simulation Parameters}
\label{Table:SimulationParameter}
	\footnotesize
\renewcommand{\arraystretch}{0.8}
	\begin{tabular}{ll}\toprule
{\bf Simulation Parameters} & {\bf Values Considered} \\\midrule
        Propagation Model & Shadowing (Log-Normal)\\
        Standard Deviation & 14 dB\\
        Deployment Area  & 200m x 200m\\
    		Transmission Range & 30m\\
		Data Rate & 250 kb/s\\
		Packet Length & 100 B\\
		Slotframe Length & 100 slots\\
		 Time slot duration & 10ms\\
       Output buffer size & 10 packets\\
       $I_{min}$ & 3s\\
       $\theta_{th}$  & 0.5\\
       $\alpha$  & 0.5\\
       $\delta_{th}$  & 0.5\\
		\bottomrule
	\end{tabular}
\end{table}

\subsection{Throughput}
Throughput is one of the most important QoS parameters for any IoT application. It can be observed from Fig~\ref{fig:EWQOF_thr} that the proposed mechanism outperforms the Max QOF technique. This is largely due to the fact that the parent nodes are not swap immediately after a sudden spike in traffic, which otherwise results in packet loss. Unnecessary swapping increases network overhead from the association process and reduces throughput. Additionally, the proposed mechanism computes an optimal parent for data transmission in a congested channel resulting in higher successfully transmitted packets. We observe a 6-30\% improvement in the RPL performance on implementing the proposed mechanism.

\subsection{Packet Delivery Ratio}
The performance metric known as the packet delivery ratio (PDR) is commonly used in sensor network literature and is a vital parameter to measure the performance of a routing protocol like RPL. It measures the proportion of successfully delivered packets to the total number of packets transmitted by the sensors. Fig\ref{fig:EWQOF_pdr}) shows that on scaling up the network, the PDR of the proposed mechanism is higher than the Max QOF technique. This can be attributed to its inherent nature of identifying and swapping parents based on the average level of congestion in the parent, $\beta_{\text{prop}}$ and avoiding needless swaps. We observe a 4-10\% improvement in the RPL performance on implementing the proposed mechanism.

\subsection{Average number of swaps}
Swapping are associated with additional overhead as discussed earlier. Frequent unnecessary swapping result in
excessive power consumption with the exchange of control
frames and transmission of messages. Fig.~\ref{fig:EWQOF_swaps} shows the total number of swaps over a fixed period of time. Max QoF suffers from excessive swaps as it selects a new parent whenever the current data burst increases beyond $\theta_{th}$. However, the proposed EWQOF mechanism has lesser number of swaps compared to the Max QOF, especially on increasing the network size beyond 30 nodes (15-60\%). 
%This is because as the network size increases the herding problem gets pronounced and problematic in networks with sudden spikes of data. Thus for larger network EWQOF performs significantly better in congestion control without making the network unstable.
%EWQOF mainly focuses on this aspect of the congestion control. Here as the number of nodes increases e.g 50 nodes EWQOF is performing significantly better. This is due to the fact more the number of nodes more the possibility of swapping since congestion can lead to herding problem. Herding problem is especially pronounced and problematic in networks with sudden spikes of data. Thus for larger network EWQOF performs significantly better in congestion control without making the network unstable.

\subsection{Average Energy Consumption}
As the devices considered in a 6TiSCH-based IoT network application are energy constrained, it is of utmost importance to have techniques that increase the overall network lifetime. The incorporation of the proposed mechanism in RPL results in higher energy savings compared to the Max QoF technique. The primary reason is the reduced number of parent swaps and the ability to compute and select the appropriate parent for the transmission route during network congestion.   
%This is the main achievement that EWQOF flaunts. With less number of swaps, or not needing to swap further enforces that less number of control signals are required. Therefore saving overall energy. 
The energy savings is further noticeable for large-sized networks as shown in Fig\ref{fig:energy}. 
%Four control signals have been considered as shown in Fig\ref{fig:model} as our measure for energy consumption during a parent swap. Additionally, energy consumed during data transmission/reception is  one packet is same as \cite{energypaper}

\section{Conclusion \& Future Work}
\label{conclusion}
In this paper, we address the issue of inefficient congestion control mechanisms in RPL arising from inefficient in-built congestion control mechanisms. The severity of the issue is observed in bursty data traffic resulting in frequent changes in the parent-child associations. In this paper, we address this issue by reducing unnecessary parent changes and computing an appropriate parent and route. This ensures that a node forwards its packets to the destination through the least congested path, resulting in minimal packet loss. A novel routing metric incorporating exponential weighting has been proposed to select the parent node based on the number of packets in the node's queue at a given moment. Additionally, the proposed parent selection and swapping mechanism enables efficient transmissions in congested networks.
The simulation results show that the proposed mechanism outperforms the related technique by reducing the number of parent swaps and energy dissipation and improving the packet delivery rate and throughput.
As a part of future work, we plan to investigate and further analyze the parameters affecting the parent swaps and implement them in a real testbed.

%For this, a new metric for routing using the concept of exponential weighting has been proposed which takes the number of packets present in the queue of the node into account when choosing the parent at a particular instance of time

\bibliographystyle{IEEEtran}
%\bibliography{nik_ref} 
% Generated by IEEEtran.bst, version: 1.14 (2015/08/26)

\end{document}